# Experimental reasoning in process tracing: a method for calculating $p$-values for qualitative causal inference

Matias López[*] and Jake Bowers[†]

2025-05-05

We introduce a method for calculating $p$-values to test causal hypotheses in qualitative research *a la* process tracing. As in an experiment, our $p$-value tells us how often one would make the same or more compelling observations favoring one theory while entertaining a rival theory. We adapt Fisher's (1935) randomization-based urn model to the reality of qualitative researchers, who cannot randomize history, but can make observations about historical processes. Our test includes a method of sensitivity analysis which allows researchers to account for the possibility of observation bias, as well as a framework for representing the varying strenght of individual pieces of evidence, altoguether informing the robustness of qualitative causal infernce. We provide simulations and replications of previously published work to illustrate how to execute our test using any type of qualitative data about events that took place within one case. This approach adds to the pluralistic turn in the use of probability theory in theory-testing process tracing by offering a simple model with provable conservatism, while relying on few assumptions the consequences of which can be directly assessed.

# 1 Introduction

The recent probabilistic turn in process tracing has made it easier to evaluate the probative weight of qualitative evidence given causal theories about single cases, with Bayesian models catalyzing a renewed interest on the uses of probability theory in qualitative research (Bennett 2008; Fairfield and Charman 2022; Humphreys and Jacobs 2015; for a critique see Zaks 2021, 2022). In this paper, we enlarge the probabilistic toolkit available to researchers by introducing a way to use *p*-values to test causal arguments when $N = 1$. A *p*-value can help us test a theory by providing a sense of the probability of observing the evidence in hand if indeed a rival theory is correct.

---

[*]Universidad Diego Portales & Albert Hirschman Centre on Democracy Geneva Graduate Institute, Escuela de Ciencia Política, matiaslopez.uy@gmail.com

[†]University of Illinois at Urbana-Champaign, Department of Political Science & Statistics, jwbowers@illinois.edu

We refer to theory-testing process tracing (Falleti 2015) given that much of the methodological debate about qualitative causal inference builds on this framework, but our approach fits within-case research broadly conceived. We build on Fisher's (1935) insight that a rival theory can be represented by a probability distribution, which allows us to calculate a *p*-value. Fisher used an urn model to generate this distribution, which we adapt to account for miscellaneous but in-depth qualitative information about a (presumably) causal chain of events in a single case. An urn model simulates the process of drawing items from a larger set, like balls in an urn. While Fisher justified his urn model using randomized treatment, we justify our urn model focusing on "conservatism."[1] That is, among all urn models that describe how observations support a rival theory or not, the urn model we propose makes it hardest to reject the rival theory, thus preventing false positive errors. Since process-tracing observations tend to arise via purposive search whereas experimental observations tend to arise in controlled environments, we develop a framework for sensitivity analysis for the assumptions behind this urn model and suggest a workflow in which sensitivity analysis plays a central role.

For reasons that should become clear, we call our urn model the "+1 hypergeometric model" or "+1 urn model." Together, the information produced by this test and by the associated sensitivity analysis provide a reasonable basis to reject or fail to reject a rival theory about one case relying solely on qualitative observational evidence about that case. We further address process tracers' concern with the strength of individual pieces of data by showing how to use the urn model to represent the probative weight of one or more observations.

Although our test (as others before) will not solve process tracing's general inability to use research design to justify a simple stochastic model (as a randomized experiment would), it builds on the rationale of randomized experiments, following what Morton and Williams (2010) called "experimental reasoning." Our use of probabilities is also inspired by Glynn and Ichino (2015) who rely on qualitative information to calibrate *p*-values, as well as on the more general probabilistic turn in process tracing led by Fairfield and Charman (2022) and Humphreys and Jacobs (2015). We add to that work by building on Fisher rather than Bayes, and in the end develop a particularly simple and probably conservative way to apply probability theory in hypothesis testing in qualitative research. By adopting *p*-values to judge the evidentiary strength of qualitative evidence, we bring the complex world of estimating the causes of effects via process tracing closer to the simpler world of testing hypotheses about the effects of causes.[2]

---

[1] Fisher's model represents the the variation in a hypothetical null relationship by showing how a result would vary across the randomizations implied by the experimental design.

[2] See Waldner (August 2025) for a recent in-depth discussion about qualitative causal inference and for an introduction to counterfactual causal inference see "10 Things to Know About Causal Inference" " by Macartan Humphreys available at: https://egap.org/resource/10-things-to-know-about-causal-inference/

Although we deviate from process tracing's traditional reliance on set theory (Barrenechea and Mahoney 2019; Beach and Pedersen 2019), we discuss how our probabilistic test relates to traditional tests, such as the smoking gun test (Collier 2011; Van Evera 1997). Meanwhile, we build on previous probabilistic approaches to process tracing by showing how foundational elements of experimental research, such as an urn model, can help us learn about counterfactual scenarios through qualitative evidence. As we will describe below, our test is attractive because it uses relatively few assumptions (with which we engage and allow researchers to relax), involves simple steps, and provides a summary that can be interpreted directly. The math behind our model is simple, and qualitative scholars can easily produce *p*-values based on their data using a few lines of R code, our own R package, artificial intelligence, or pen and paper, as we show in the supplementary materials.

In what follows, we first describe the idea of counterfactual causality and the problem of causal inference in $N = 1$ studies. We then explain our method in detail using simulations and apply our approach to a recent study by Rossel et al. (2023) explaining a shift in Uruguay's conditional cash transfers policy. We also discuss the limitations of our test by comparing *p*-values produced by our model with those produced by an actually randomized experiment: Fisher's (1935) famous "Lady drinking tea" problem.

## 2 Setting the problem

Process tracing, broadly defined, is a method for using observations to evaluate whether/how a cause is connected to one outcome within one case (Falleti 2015; Van Evera 1997; George and Bennett 2005). This evaluation may take the form of "theory building" or "theory testing" versions of process tracing, or a combination of these (Falleti 2015). We focus on the "theory testing" approach in this paper. The terms "case" and "observation" are often used interchangeably in research (Hall 2013). We think of a case as a unit within a broader population of cases (Gerring 2004). For example, the French revolution can be one case within the population of revolutions.[3] Meanwhile, we use "observations" to refer to data relevant to a case. Observations relevant to the French revolution might include a speech by Robespierre or the public tax accounts published in 1781.

Often researchers want to find the cause of a single outcome in one case, even if they mobilize diverse observations to support or oppose one or more causal explanations. While process tracing does not exclude the analysis of multiple cases, the method implies the separate "tracing" of causal processes within each case, rather than entertaining causality by matching cases as in the comparative history[4] tradition (see T. Skocpol and M. Somers 1980; Thelen and Mahoney 2015; also Ragin 1999). The focus on individual cases implies that there might be no relevant source of variation in the outcome of interest itself, but there can be quite a lot of variation in the information that we are able to collect about this outcome. Thus, lack of variation is not necessarily the main problem regarding causal inference about one case.

---

[3] The French revolution could also be thought of as one case within the population of things that happened in France, or one case of events in 1789, etc.

[4] Other related methods are Comparative Historical Analysis and Qualitative Comparative Analysis, which focus on commonalities across cases.

The "fundamental problem of causal inference" (Holland 1986) is not about how many observations or cases one has in hand (see also Titiunik 2015). It is about the impossibility of knowing for sure that some explanatory variable or treatment, $T_i$, caused some outcome, $y_i$, for case $i$. Such certainty could only come from observing both case $i$'s response when $y_{i,T=1}$ and when $y_{i,T=0}$.[5] For example, to *know* the causal effect of the publication of the tax accounts in 1781 on the launch of the French Revolution, we would need to observe two versions of France: one where, as we know, the tax accounts were published, and another counterfactual France, in which the tax accounts were not published. Since observing both versions of France is impossible, instead we can try to *infer* about the counterfactual France in order to make a counterfactual causal inference (see Imbens and Rubin 2015).

Researchers use different criteria to assess what observations say about counterfactuals, often in the form of a hypothesis test. Process tracers traditionally work with the idea that a given hypothesis, $H_T$, may definitively pass a hypothesis test if observations seem logically compatible with $H_T$ alone (as in "smoking gun" and "double decisive" tests), or they may drop $H_T$ if logical derivations of $H_T$ are not observed (as in the "hoop" test).[6] Such an approach borrows from deterministic ideas about sufficiency and necessity in set theory (Beach and Pedersen 2019). To illustrate, let us call *observation 1* a "smoking gun," a piece of evidence that we see as sufficient for confirming that $T$ is the cause of $y$ within country $a$. The sense of certainty that *observation 1* may produce contrasts with the fundamental problem of causal inference, which tells us that we cannot know for sure that $y_{a,T=1} > y_{a,T=0}$ is true because we cannot observe $y_{a,T=0}$. Instead we can say that *observation 1* seems very odd unless $T$ is the cause of $y$, and this raises the question: How odd? How can we have a sense of the probability of making *observation 1* if the reality is that $T$ was *not* the cause of $y$?

Experimental research has long offered a simple answer to this question using *p*-values, developed originally in Fisher's "Lady Tasting Tea" problem (Fisher 1935, Chapter 2). But we cannot apply this solution directly in process tracing because Fisher's *p*-value arises from a probability distribution that is justified by a randomized treatment.

To recall the history of the *p*-value: Fisher's colleague, Dr. Muriel Bristol, declared "that by tasting a cup of tea made with milk she can discriminate whether the milk or the tea infusion was first added to the cup" (11). Fisher addressed her claim with an experiment: he prepared 8 cups of tea with milk or tea poured first at random and asked Dr. Bristol to identify the 4 cups with milk added first. Fisher's test focused on a counterargument to Bristol's theory, a null hypothesis, according to which Dr. Bristol ignores the flavor of the cups of tea and, in essence, has already decided on her answers for each cup before tasting any of them such that the flavor of the cups can have no effect on her tasting decisions. The story goes that Dr. Bristol arranged the cups correctly. Fisher showed that in this experiment, the probability of getting all cups right by chance is exactly $P(\text{number correct} = 4) = \left(\frac{8!}{4!(8-4)!}\right)^{-1} = 1/70 = 0.014$.[7] Thus, in a world where Dr. Bristol guesses, one would observe this perfect run in only 1.4% of all possible trials. This probability became known as the *p*-value, and we can use a low *p*-value to conclude that a rival theory is most likely incorrect.[8]

---

[5]To learn about an additive causal effect we would need to calculate $y_{i,T=1} - y_{i,T=0}$.

[6]For definitions and a guide of how to implement Van Evera's (1997) tests see Collier (2011).

Whereas Fisher made it clear how we can use a randomized treatment to calculate a *p*-value and use it to infer about counterfactual scenarios, in process tracing one must do something different to generate a distribution that represents a rival theory. This is the key problem that we aim to solve.

## 3 The urn model without randomization

In Fisher's tea study, the null hypothesis states that Dr. Bristol would not change her guesses regardless of the order of milk and tea. In the literature on the statistics of causal inference we might write, $H_0 : y_{i,T_i=1} = y_{i,T_i=0}, \forall i$ to represent the idea that she would have made the same answer for each cup $i$ regardless of whether milk was added first, $T_i = 1$, or second, $T_i = 0$. The null distribution represents the ways that Bristol might make choices consistent with this hypothesis arising from the rival explanation across different randomized orderings of the cups. In process tracing, if we wish to test the theory $H_T$: $T$ caused the event $y$ in country $a$, it may seem unreasonable to assume that $T$ was assigned at random to case $a$. But we do not have to make that assumption to formalize a counterargument to $H_T$, we can use social science theory and then rely on other assumptions to build a probability distribution.

Imagine that the researcher has observations from 2 interviews which account for presumably strong evidence in favor of the argument that $T$ caused $y$ in $a$ because that same evidence seems at odds with other explanations. To test $H_T$, the researcher wishes to calculate just how probable it would be for her to make the same observations if $H_T$ was wrong. This task requires us to specify a rival explanation, one that antagonizes $H_T$. Thus we must entertain the idea that another causal driver, $R$, caused the event $y$, implying that $y$ would have occurred in country $a$ regardless of $T$, all else equal. Let us call the rival hypothesis $H_R$. If $H_R$ is correct, it follows that $T$ had no effect on $y$ within case $a$, or simply $H_R : y_{a,T=1} = y_{a,T=0}$.

The rival or null hypothesis $H_R$ does not provide information about probability on its own. To calculate a *p*-value as a measure of evidence, we need a probability distribution that represents $H_R$ but also fits the observed evidence favoring $H_T$. We develop this distribution using a simple model in which a researcher can make different types of observations. One type of observation provides evidence that favors $H_T$ and not $H_R$. In our running example the researcher made two of such observations via interviews. Yet we must entertain the possibility of making a second type of observation, one that is either in favor of $H_R$ and against $H_T$, or compatible with both $H_T$ and $H_R$. Let us divide these two types of observations in two sets. In set $\mathcal{T}$ we have the observations favoring $H_T$ alone and in set $\mathcal{R}$ we have the observations that seem compatible also with $H_R$.

---

[7] We also show how to calculate this probability using direct simulations from an urn-model in the supplementary materials. Fisher derives the closed-form expression in the 15 pages of his 1935 textbook.

[8] Underlying Fisher's approach is the idea that what we observe is fixed — Bristol did guess correctly 4 times — whereas what we do not observe is all of the ways that the tea evaluation might have occurred if, in fact, milk-vs-tea order was irrelevant to Bristol. In this approach, we have only a single probability distribution, this is why Fisher used a probability model to describe the null or rival theory rather than the observed data.

Now imagine an urn $U$ containing the two sets of types of observations such that the total size of the urn is $U = |\mathcal{T}| + |\mathcal{R}|$, where $|\mathcal{T}|$ and $|\mathcal{R}|$ represent the size of each set. And imagine drawing two pieces of evidence from this urn multiple times to represent the different ways that the research could have turned out. Sometimes both observations should support the rival theory and sometimes they might both support the working theory.

We would like for the null model to represent a world where the rival theory is true (for the sake of argument), therefore we must calibrate the urn so that $|\mathcal{R}| > |\mathcal{T}|$ because the expected evidence must favor the null hypothesis. In this example, $\mathcal{R}$ must contain at least $|\mathcal{T}| + 1 = 2 + 1 = 3$ pieces of information, even though such evidence was not actually observed. To motivate the urn model supporting $H_R$, one could speculate that the researcher missed the evidence from other interviews or archival documents which would have suggested directly that $H_R$ was a better explanation than $H_T$.

So let us work with an urn containing 2 pieces of information from $\mathcal{T}$ (which were directly observed) and 3 pieces from $\mathcal{R}$, so that $U = 5$. In the next section we show why assuming $|\mathcal{R}| = |\mathcal{T}| + 1$ provides the most conservative value of $p$. For now let us focus on the fact that this urn represents the idea that only 2/5 of the data were actually observed. Let us also make the simplifying assumption that all 5 observations are equally likely to be made. We will show later how to relax this assumption.

How many ways can we draw 2 items from a set of size 5? Let us label the items 1 to 5 and imagine that items labeled 1 and 2 are the interviews supporting $H_T$. There are 12 possibilities to draw 5 items where items 1 and 2 are either the first or the second choice.[9] But there are 5!, or $5 \cdot 4 \cdot 3 \cdot 2 \cdot 1 = 120$ possible ways to choose 5 items from the urn in any order. Since there are 120 ways to draw 5 items from the urn, but only 12 ways to choose 2 items supporting $H_T$, the $p$-value which summarizes the evidence against the rival is 12/120 or $p = .10$.

Fisher knew the exact shape of the probability distribution that describes all of the possible ways for the experiment to produce results consistent with the null hypothesis, and thus Fisher's test and $p$-value are known as "exact." Without randomization, but with this simplified model of the process of making observations, we know an exact upper bound on the $p$-values. So when this model generates $p = .1$ this actually means $p \leq .1$. For this reason we will use inequalities to communicate the value of $p$.

In Figure 1 we locate the observed data in the null distribution. The figure shows that at least 90% of the time one would make either 0 or 1 observation favoring $H_T$ in this distribution. But $p$ represents the probability of observing what was observed. The researcher would make the strong observations from interviews 10% of the time. This $p$-value suggests relatively strong evidence against the rival.

---

[9] We show this graphically in the online supplement.

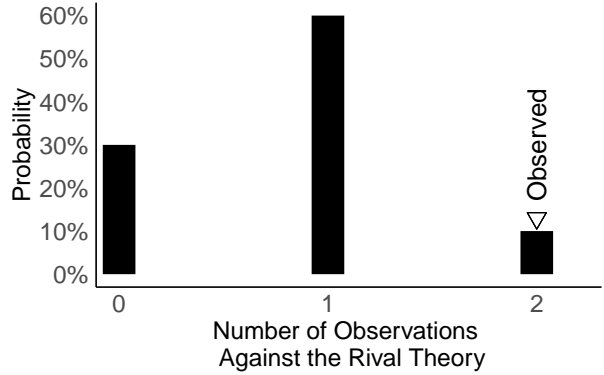

Figure 1: Null distribution for positive evidence from 2 interviews

In summary, this urn model allows us to calculate the probability of making $x$ observations from $\mathcal{T}$ (in favor of $H_T$) within a sample of size $n$, assuming that the urn contains items from both $\mathcal{T}$ and $\mathcal{R}$, and that the total number of items in the urn is $U = |\mathcal{R}| + |\mathcal{T}| = |\mathcal{T}| + 1 + |\mathcal{T}|$, or simply $U = (2|\mathcal{T}|) + 1$. The mathematical expression for this probability is:

$$P(x) \leq \frac{\binom{|\mathcal{T}|}{x}\binom{U-|\mathcal{T}|}{n-x}}{\binom{U}{n}}. \tag{1}$$

If we apply this formula to our running example, we see that it produces the same answer that we would get if we were to count up the possibilities by hand: $P(x=2) \leq \frac{\binom{2}{2}\binom{5-2}{2-2}}{\binom{5}{2}} = \frac{1}{10} = .10$.

What Equation 1 describes is the density function for the hypergeometric probability distribution, which is one of the simplest models for drawing items from an urn containing only two types of items. It is the same function used by Fisher in the tea problem. We call our version the "+1 hypergeometric model" to emphasize two simple assumptions: (i) that the discrete distribution of observations barely favors the rival by the margin of one additional observation, and (ii) that each piece of evidence carries the same inferential weight or relevance. As unrealistic as the first assumption is, it makes for a more conservative summary of evidence against the rival, as we show next. The second unrealistic assumption can be relaxed to produce either more or less conservative results as we show in Section 4.

## 3.1 Why this $p$-value is conservative

If the goal is to represent a world where a rival theory is true, why did we allow $\mathcal{R}$ to have only one more observation than $\mathcal{T}$?[10] We show here that this unrealistic assumption privileges Type II Error, i.e. our calibration of $\mathcal{R}$ makes the probability of failing to reject the rival theory when it is false more likely than the error of rejecting the rival when true. Because the model is prone to err on the side of supporting the rival, it avoids false positives. This is why we call our approach conservative.

Figure 2 illustrates this conservativeness for the simulation of a study with observations from 2 interviews supporting $H_T$. It shows that the more elements from $\mathcal{R}$ we place in the urn, the lower the probability of making these 2 observations, the more deflated are the resulting $p$-values, and the stronger the evidence against $H_R$. We prefer the opposite, i.e. to give the benefit of the doubt to $H_R$.

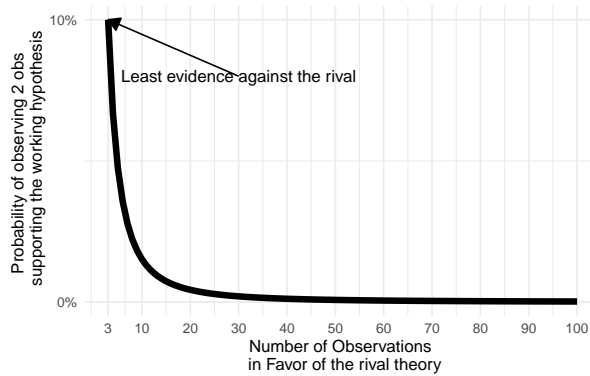

Figure 2: Probability of drawing 2 positive observations in urns with difference sizes.

In the running example, with $|\mathcal{R}| = |\mathcal{T}| + 1 = 2 + 1 = 3$, we have $p \leq .1$. But with $|\mathcal{R}| = 10$ we would have $p \leq .015$ and with $|\mathcal{R}| = 100$, we would have $p \leq .0002$. Any discrete value of $|\mathcal{R}|$ greater than 1 produces a larger $p$. Meanwhile, a discrete value smaller than 1, i.e. 0, breaks the null model because with $|\mathcal{R}| = 0$ it is no longer true that the average or expected evidence favors $H_R$ in the model. In a nutshell, $|\mathcal{R}| = |\mathcal{T}| + 1$ provides the most conservative $p$-value in a world giving the benefit of the doubt to the rival theory, where $|\mathcal{R}| > |\mathcal{T}|$.

We can also see this conservatism in action by referring to the formula for hypergeometric distribution, which we write below in Equation 2 using $|\mathcal{R}| + |\mathcal{T}|$ instead of $U$.

$$P(x) \leq \frac{\binom{|\mathcal{T}|}{x}\binom{(|\mathcal{R}|+|\mathcal{T}|)-|\mathcal{T}|}{n-x}}{\binom{|\mathcal{R}|+|\mathcal{T}|}{n}}. \tag{2}$$

---

[10]We discuss in the online supplement what to do in the rare event that most observations already favor the rival.

If $|\mathcal{T}|$ is fixed (given that we actually made 2 observations), and if our model involves choosing $n = 2$ observations at a time to match the actual observations, then the probability of observing $x = 2$ declines as the denominator increases (adding observations to $\mathcal{R}$). When $|\mathcal{R}| = |\mathcal{T}| + 1$, $\binom{|\mathcal{R}|+|\mathcal{T}|}{n}$ is at its smallest while still giving the benefit of the doubt to the rival theory.

This discussion is the intuition behind the following theorem of conservativeness:

**Theorem 1.** *Consider the tail probability of the cumulative density function of the hypergeometric distribution:*

$$F(x) = P(X \leq x) = \sum_{j=0}^{x} \frac{\binom{|\mathcal{T}|}{j}\binom{|\mathcal{T}|+c}{n-j}}{\binom{2|\mathcal{T}|+c}{n}}, \tag{3}$$

*where we write $|\mathcal{R}| = |\mathcal{T}| + c$, and where $0 \leq x \leq n \leq |\mathcal{T}|$, $c \geq 1$, $n = |\mathcal{T}|$, and $|\mathcal{T}|, n, x, c$ are nonnegative integers.*

*The theorem states that, for fixed $x, n, |\mathcal{T}|$ the tail probability, $1 - F(x) = P(X > x) = \sum_{j=x+1}^{n} \frac{\binom{|\mathcal{T}|}{j}\binom{|\mathcal{T}|+c}{n-j}}{\binom{2|\mathcal{T}|+c}{n}}$, is maximized when $c = 1$.*

We prove the theorem in the in the supplementary materials. Yet another aspect of this urn model makes for a conservative test. Notice that, at this stage in the development of the +1 urn model, we assume that each observation carries equal inferential weight although we know that in reality some pieces of evidence may be much more compelling than others. We do not need to make this assumption and we show how to account for evidentiary weight in Section 4. However, by equalizing the relevance of each piece of evidence, we keep the calculation of $p$ unaffected by "silver bullets," or single pieces of evidence that strongly corroborate the working hypothesis. Once more, this decision at least potentially inflates the value of $p$ in light of some very compelling observations against the rival, reinforcing the general conservativeness of the test. Such conservativeness is completely independent from the type of data used in the research, be they interviews, documents, ethnographic notes or any mix of qualitative information. In the running example, the researcher has two interviews that account for $p \leq .1$. In the next sections we show a series of steps that she could take to learn more from $p$-values.

### 3.2 Adding observations

If the researcher wishes to learn more and collects further information about event $y$ in country $a$, the $p$-values that the urn-model generates should reflect that. Assume that the researcher now has a dataset of 7 observations from interviews and documents supporting her theory and 3 observations from similar sources supporting the rival, with a new total of 10 observations. The urn would now contain $U = |\mathcal{R}| + |\mathcal{T}| = 2|\mathcal{T}| + 1 = 15$ items, accounting for the 7 observations assigned to $\mathcal{T}$, all of which were actually made, and $|\mathcal{T}|+1=8$ observations in $\mathcal{R}$, of which only 3 were actually made.

Equation 1 allows us to calculate the probability of observing as many pieces of evidence supporting $H_T$ out of total of 10 observations drawn from the urn. Figure 3 shows that it would be much rarer to make 7 observations supporting $H_T$. The urn model predicts that, if the data collection process were repeated with 10 observations made each time, the researcher would make the same or more compelling observations only 1.9% of the time tops, because $p \leq 0.019$.

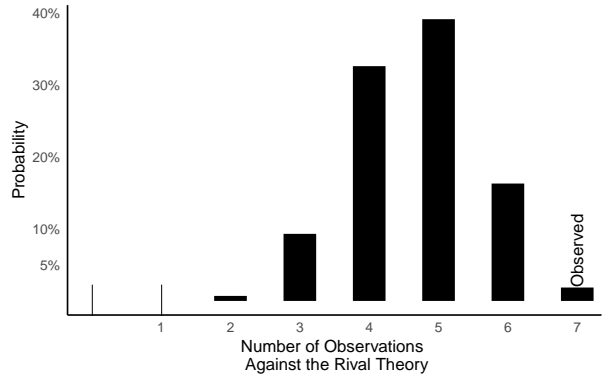

Figure 3: Null distribution for positive evidence from 7 out of 10 observations.

Of course, although it would be very *improbable* to see so many observations supporting the working theory under the rival hypothesis, these observations must remain *possible* under the null. This is part of the logic of hypothesis testing. However, one can say with confidence that such observations would be *unlikely* under $H_R$ and one can also provide a measure for this probability, $p \leq .019$. The researcher can report that the data cast great doubt on $H_R$.

This *p*-value depends on two assumptions: that there was no significant difference in odds of observing one or another of the pieces of evidence, and that observations carry the same evidentiary weight. The conclusion that the data go against $H_R$ may thus be sensitive to whether these assumptions hold, or it may not. To learn about sensitivity, we propose a framework for sensitivity analysis as part of the workflow of using this test.

## 4 Sensitivity analysis

In case study research, we should not be surprised if certain kinds of evidence are harder to come by than others. Maybe interview subjects have reason to downplay evidence that supports the rival theory, maybe strategic actors have destroyed evidence before it made its way to the archive. If such difficulty is correlated with the type of information that data bring, this causes bias. If bias in observations were present, it may be that our *p*-value upper bound is overly optimistic about how much evidence there is against a given rival theory. We may never know if bias was indeed relevant, but we can know if our substantive interpretation of results would differ if the observations made were overly likely to support $H_T$. A biased urn model provides this information. By "biased" we do not mean that this is somehow a worse urn. Rather, a "biased urn" is a term of art for an urn model where certain kinds of items are more likely to be drawn than others (Fog 2023).

The question of whether bias was or was not present is less relevant for the test. Most observational research designs have some observation biases. Certainly this is the case for most data collection processes in qualitative case study research, as researchers are not randomly sampling all potential sources of information from some known population of sources. But the conclusions of a study based on qualitative data may be relatively insensitive to a degree of such bias when it comes to the interpretation of the results, or they may not. A sensitivity analysis does not reveal bias. It is a formalized thought experiment about bias, and whether it would change our conclusions.

We can account for different odds by allowing the probability of making one type of observation to be governed by the parameter $\omega$.[11] We write $|\mathcal{T}|$ and $|\mathcal{R}|$ (rather than using $U - |\mathcal{T}|$ for $|\mathcal{R}|$) to make it more clear that we are summing up the possible ways to draw items. This model, shown in Equation 4, allows for different odds of drawing a particular kind of observation from the urn.

$$P(X = x) \leq \frac{\binom{|\mathcal{T}|}{x}\binom{|\mathcal{R}|}{n-x}\omega^x}{\sum_{j=\max(0,n-|\mathcal{R}|)}^{\min(n,|\mathcal{T}|)} \binom{|\mathcal{T}|}{j}\binom{|\mathcal{R}|}{n-j}\omega^j}. \tag{4}$$

Imagine a researcher calculates a low $p$-value upper bound using the basic urn model (say, $p = .01$). If it would require a large bias ($\omega$) to change the $p$ to, say, above .05, then we interpret the results as relatively insensitive to bias. To know how much bias is needed to reach a certain threshold, we can hold fixed the $p$-value and solve for $\omega$.

We can find $\omega$ such that $p(x) \equiv P(|\mathcal{T}| = x) = \alpha$, where $\alpha$ is some threshold beyond which we claim that the observed data casts a great deal of doubt on the rival.[12] To provide a sense for how $\omega$ relates to $p$, we show the solution for the simulation of a researcher with a dataset of 10 observations from interviews and archives, among which 7 favor $H_T$. To recall, our test provided $p \leq .019$ in that case. While this suggests the rejection of the null, the researcher may question if bias led her to this conclusion. She cannot know if there was bias, but she can ask: How much observation bias would be needed to inflate this $p$-value up to .05 or .10? Equation 4 sets the $p$-value for this model as a function of $\omega$ as follows:

---

[11] In this first paper on this topic we restrict attention to the case of 2 types of information. This is another limitation of the method at this writing, but is not inherent in the use of urn models in general.

[12] This $\alpha$ can also be understood as a rejection threshold for the test.

$$P(X = 7) \equiv p \leq \frac{\binom{7}{7}\binom{8}{10-7}\omega^7}{\sum_{j=2}^{7}\binom{7}{j}\binom{8}{10-j}\omega^j} = \frac{8\omega^5}{3 + 40\omega + 140\omega^2 + 168\omega^3 + 70\omega^4 + 8\omega^5}. \quad (5)$$

While solving quintic polynomial equations algebraically is difficult, finding $\omega$ for a specific $p$-value numerically is quite easy, and we display the R code we used in the supplementary materials. Figure 4 shows how the $p$-value from equation 5 rises as bias increases, eventually reaching $p \leq .05$ and $p \leq .10$ levels. In our simulation, one would need $\omega = 1.58$ for our calculated upper bound to include $p = .05$ or $\omega = 2.34$ for $p = .10$. These are odds ratios and they can be interpreted directly: 1/1 entails equal odds, thus 0% change necessary to flip conclusions. In this simulation, a ratio of 1.58 suggests that observations favoring $H_T$ would have to be roughly 60% more likely in order to change the $p$-value from .02 to .05. If the researcher adopted a $p < .1$ threshold, $\omega$ indicates that favorable observations would have to be 130% more likely to be made in order for bias to invalidate her substantive conclusion. That is, the substantive interpretation of results would remain unaffected up to a point of $\omega = 1.58$ or $\omega = 2.34$, depending on which threshold the researcher wants to use to reject the rival. The higher the odds-ratio, the lower the sensitivity.[13]

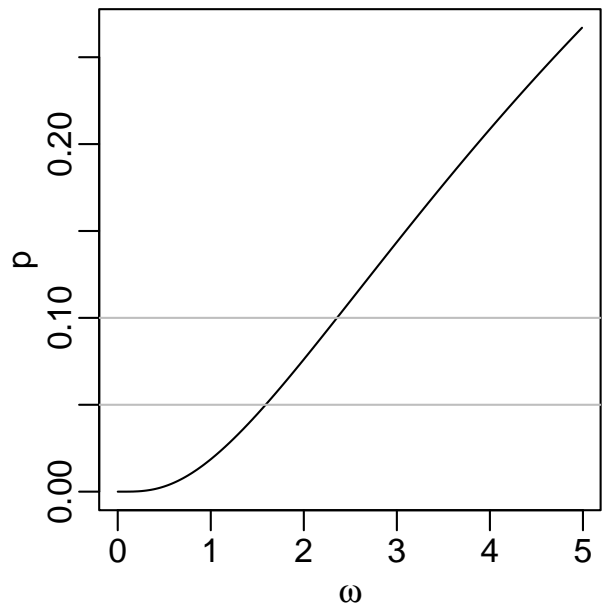

Figure 4: P-values as a function of $\omega$ for 7 positive observations out of 10.

In this simulation, we have a low $p$-value and low sensitivity. Knowing this, the researcher can now more confidently announce a positive result, even considering that maybe the archive did not store some important information about case $a$, or that maybe the interviewees who were willing to talk were also likely to mildly downplay the role of the rival explanation about case $a$.

---

[13] In our supplementary materials we show how to find closed form solutions for simpler versions of this equation to verify the numerical solutions from R.

# 5 Accounting for evidentiary weight

A related issue concerns the weight that some observations may have in favor of one theory or the other. Scholars of process tracing often highlight how a sufficiently strong observation can operate as a "smoking gun" or a "double decisive" test, in effect providing conclusive insight about causal theories (Collier 2011). Weighting the evidence to distinguish more conclusive pieces is not demanded by our test, and this practice certainly opens a discussion about which protocols are best suited for discriminating the strength of evidence. Be it as it may, if a researcher wishes to weigh her data, doing so in the urn model is straightforward.

Imagine, for example, that we have 4 observations, 3 of which clearly support the working theory $H_T$ while 1 clearly supports the rival. The +1 urn model would set $|\mathcal{T}| = 3$ and $|\mathcal{R}| = 4$, producing $p \leq .11$, implying a probability upper bound of 11% of observing the same or more compelling evidence if $H_R$, not $H_T$, was correct. Now let us assume that the first of such pieces of evidence is particularly strong, a "smoking gun" strongly indicating that $T$ caused $y$ in case $a$. One can entertain several ways of weighing this individual piece of evidence. Fairfield and Charman (2022) suggest we use decibels as an intuitive way to think about weight of evidence and we follow them in this regard. So imagine that two pieces of information "speak" in favor of the $H_T$ while that one observation "screams" that $H_T$ is true. In decibels this implies a ratio of about 60 to 120, making that one observation roughly twice the value of the other pieces of evidence. What would this imply?

If observation 1 is worth 2 other observations, the urn can reflect this by increasing the size of $|\mathcal{R}|$, accounting for the weight of that smoking gun observation. So now $|\mathcal{R}| = (|\mathcal{T}| + 1) + 1$. The latter calibrates the urn accounting for the weight of one observation without implying that we have more observations than we actually do. That is, we will still be sampling 4 observations from the urn. Of these, 1 speaks against the researcher's theory, 2 speak in favor of the theory, and 1 screams that the theory is true.

This new urn model with one up-weighted working-theory observation yields $p \leq .071$ instead of $p \leq .11$. The three observations are stronger evidence against the rival theory. The test now tells us that it would be even more improbable to make these 3 observations, or even stronger observations, given the rival model. This new summary is more conclusive. But what if these three observations were particularly easy to observe because of observation bias? Figure 5 shows the probability space generated by different weights for the first observation and with different levels of bias (represented by $\omega$). We mark the $p \leq .071$ location, representing the evidence against the rival if we consider an evidence weight of 2 for one smoking gun observation and the assumption that rival-supporting and working-theory-supporting observations are equally likely from this urn. At the very bottom we have $p \leq .11$ representing an urn with no individual observation worth more or less than any other. At the top we also see $p \leq .11$, even if we had an incredibly powerful single smoking gun (with 5 times the weight of any other observation), but such observations were 2.5 time more likely to be observed given the state of the archives or interests of the interview subjects. In that scenario we might fail to reject the rival, even if one observation is a "smoking gun."

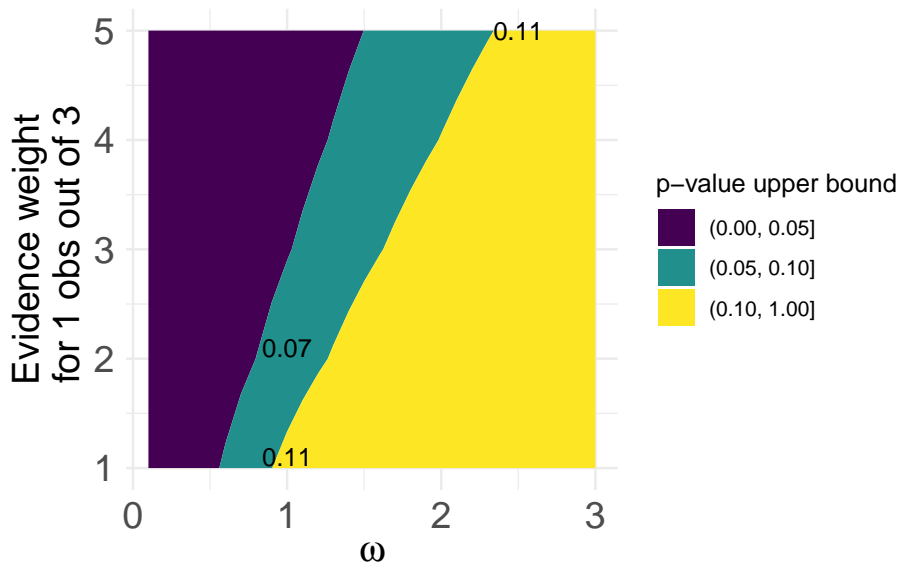

Figure 5: *p*-value upper bound areas according to weights and $\omega$.

Yet, this plot also shows that any combination of observation bias or evidence weight in the green or purple regions of the plot would yield strong evidence against the rival (with $p < .1$).

In sum, our method allows researchers to easily represent the idea that some observations carry more inferential weight than others. This can be done using the decibels metaphor or by any other criterion. Meanwhile, the probability space that these decisions generate allow researchers to understand how dependent their conclusions are of assumptions about the weight of evidence and how such assumptions interact with other assumptions about the context and process of making the observations themselves, and this is useful for communicating the conclusiveness of research.

# 6 Dealing with multiple rival hypotheses

For the sake of simplicity we have shown simulations with only one rival theory, whereas the reality of qualitative research often engages multiple rival theories at once. Our test extends naturally to this circumstance: the same procedure used for a first rival $H_{R1}$ can be repeated for $H_{R2}$, $H_{R3}$, and so on. This means that resulting *p*-values will express the probability of finding the same or more compelling observations in the different urn models, each representing one rival explanation or combinations of rival hypotheses. So one observation may be improbable with respect to for $H_{R1}$ but expected for $H_{R2}$ or $H_{R3}$.

Using our approach in this context may require some additional assumptions or some statistical adjustments. For example, the researcher can state hypotheses to emphasize their mutual exclusiveness, and assume that rejecting one rival does not affect the odds of rejecting a second and third rival. In a scenario with about a handful of available rival explanations, these assumptions may hold. In any case, the sensitivity analysis will continue to provide an assessment of how much different odds may affect results, and the researcher can rely on them as multiple tests are run.

If the reality of research is that of a long list of rival explanations, then these assumptions may not hold and the researcher might have to do something else to account for cumulative probabilities. For instance, she may adjust the critical value for rejecting each null, starting at $\alpha = .05$ when testing $H_{R1}$, then adjusting to $\alpha = .025$ (i.e. half) for $H_{R2}$ and so on. Another trickier scenario would be one in which hypotheses coincide or converge (Zaks 2017), either because both $T$ and $R$ independently cause $y$, or because $y$ was produced conjointly by $T$ and $R$. In that scenario we should imagine that the rival urn models are correlated, therefore testing three rivals may not be as simple as generating three independent urn models and calculating three $p$-values. One alternative would be combining the different urn models, making $\mathcal{R}$ a set of observations that favor $H_{R1}$ or $H_{R2}$ or $H_{R3}$. We foresee futher methodological development in regards to testing multiple hypotheses using urn models, but the fundamentals of our +1 model remain unaltered. All the researcher needs to do is to continue to operate with probability theory and the logic of experimental designs.

## 7 Application in the case of Uruguay's policy shift

The Frente Amplio (left-wing) administration in Uruguay implemented a conditional cash transfers policy (CCT) in 2005. By law, transfers to poor families were conditioned on school and primary health attendance, but these conditionalities were largely overlooked by government officials. In 2013, the Frente Amplio (FA) was still in power, however the government suddenly shifted to enforcing conditionalities, asking beneficiaries to prove school attendance and health maintenance or else risk exclusion from the program. In practice, this policy shift imposed sanctions on the poor. Why did the FA, a pro-poor party, shift its policy from lax to punitive?

Rossel et al.(2023) theorize that the shift occurred because the FA was seeking to appeal to middle and upper-class voters who resented "undeserving" beneficiaries and were gravitating towards the opposition. According to this theory, electoral competition was the causal driver of the policy shift.

The authors then theorize rival explanations. For instance, they entertain the idea that a change in government composition could have caused the shift. It certainly could, but we know that this change did not occur. Another rival argument states that the lack of state capacity was the main cause, and Rossel et al. tell us that indeed the government had to develop tracking tools in order to implement the shift. However, the question remains: Why did the government decided to build state capacity to shift its cash transfer policy? Among the rivals that Rossel and her colleagues entertain, one stands out as a strong candidate for the null. What if technocratic motivations were behind the shift? Maybe officials learned that a punitive policy achieved better education and health outcomes. This rival explanation also makes sense and could produce the same events observed. This explanation would also be consistent with the literature on technocrats' means tested rationale (Leão 2022). If policy learning was the true causal driver, than electoral competition could not have caused the shift.

In this example, we have two alternative processes that could have produced the same outcome and that are compatible with what we know in advance about cash transfers and about the case of Uruguay. These are:

$H_T$: Frente Amplio implements lax CCT $\to$ Opposition appeals to resented affluent voters $\to$ Frente Amplio shifts to punitive CCT.

$H_R$: Frente Amplio implements lax CCT $\to$ Officials learn that enforcement improves the policy $\to$ Frente Amplio shifts to punitive CCT.

Rossel et al.'s process tracing accounts for multiple pieces of information that favor $H_T$. In praise of simplicity, we provide a somewhat stylized version of their observations:

*Observation 1*: Opposition leaders campaigned heavily against the government's lax approach to CCT.

*Observation 2*: Polls show Frente Amplio losing ground, as well as declining support for redistribution, among more affluent voters.

*Observation 3*: Op-eds (aimed at middle-class audiences) echoed the right's pro-enforcement discourse and none of those voices came from pundits or opinion leaders that traditionally support the FA.

*Observation 4*: Public officials (FA appointed) claimed pressures by public opinion and the opposition as reasons to enforce conditionalities.

*Observation 5*: President Mujica publicly acknowledged pressures from "the middle class" against the lax policy.

*Observation 6*: FA officials unanimously stood by the lax approach before the policy shift.

*Observation 7*: Government officials held press conferences to advertise the enforcement of conditionalities after the policy shift.

*Observation 8*: Policy briefings informed officials that the positive effects of cash transfers on education and health were independent of beneficiaries' knowledge of conditionalities.

To create the urn, we code whether or not each observation clearly supports one of the two explanations. Observation 1 supports both $H_T$ and $H_R$ because one would expect the right-wing opposition to rally against the government also in the world of the null. All other observations seem unlikely to support $H_R$. For instance, observing polls showing declining support for redistribution and for the left-wing government among more affluent voters (observation 2) looks like something one would find if $H_T$ was correct and is not implied by $H_R$, thus a member of $\mathcal{T}$. Likewise, observation 3 does not seem to support $H_R$ because one would expect voices that are sympathetic to the administration to express pro-enforcement views as they learn that this was the better policy, if $H_R$ was true. It would be even stranger, from the perspective of $H_R$, to observe FA leaders acknowledging pressure from more affluent voters if indeed such pressure did not motivate them to change the policy (observations 4 and 5). Observing officials defending the lax approach as the best approach (observation 6) is also not in accordance with $H_R$. Neither is observing FA officials advertising the enforcement of law after the shift (observation 7), because this is what politicians would do if they are trying to reach voters. This is true also for another strong piece of evidence: The researchers found only policy briefings that do not suggest the need of enforcement (observation 8). This would be bizarre if one assumes that technocrats were learning the opposite. Therefore, observations 2, 3, 4, 5, 6, 7, and 8 all go to $\mathcal{T}$, making $|\mathcal{T}| = 7$. But maybe policy briefings indicating policy learning were in another archive, and interviews with better informed FA leaders were not made. These hypothetical unobserved data belong in $\mathcal{R}$. Assuming $|\mathcal{R}|=|\mathcal{T}|+1$ we now have $U = 15$. The test summary in Table 1 portrays the probability of making at least 7 (out of 8) observations supporting $H_T$ given the rival or null model.

Table 1: Test summary for Rossel et al. (2023)

| $p$-value upper bound | Odds ratio for $\omega_{\alpha=.05}$ | Odds ratio for $\omega_{\alpha=.10}$ |
|---|---|---|
| 0.001 | 4.216 | 6.292 |

The test posits that one would observe the same or more convincing evidence only $100(0.001)\% = 0.1\%$ of the time in this urn. The results have very low sensitivity, as very large biases would be needed to surpass the thresholds of $p < .05$ or $p < .1$, as expressed by odds-ratios in the second and third columns of Table 1. The odds of observing evidence favoring $H_T$ would have to be at least 4.2 or 322% more likely to be observed than rival supporting evidence to reach $p = .05$ and 6.3 or 529% more likely for an upper bound of $p = .10$. Even if the research somewhat overrepresented evidence favorable to $H_T$, the observations still cast great doubt on $H_R$. It would take very severe overrepresentation of evidence in favor of $H_T$ to weaken the evidence summarized by the $p$-values here.

Finally, some pieces of evidence are clearly stronger than others in Rossel et al's study. For instance, observing President Mujica explicitly communicating that he was under pressure from more affluent voters utterly fits $H_T$ and is something that would be very strange to find in a world governed by $H_R$. If this observation "screamed" at us, instead of "taking," (implying a weigh of 2), we would have $p \leq .0003$. The strength of this evidence would require yet more extreme differences in odds of observations supporting $H_T$ in order to overturn the idea that, while it would be possible to observe this shift in conditional transfers under the rival, it would be extremely improbable. One can confidently reject $H_R$. If we are still unconvinced that $H_T$ is correct, we can rerun the test with a second rival, maybe adjusting the critical $p$-value for rejection.

# 8 Limitations

Fisher justified his urn model using a randomized treatment. The urn model that we propose here also arises from design, but design is not enough to justify the $+1$ urn model.[14] We had to address the following three questions:

   a. How many elements from $\mathcal{R}$ should the urn contain?
   b. What is the probability of drawing a piece from $\mathcal{T}$ versus $\mathcal{R}$?
   c. What is the evidentiary weight of any given piece from $\mathcal{T}$?

To calibrate $\mathcal{R}$, we specified that $|\mathcal{R}| = |\mathcal{T}| + 1$. This answers (a) in the most conservative way possible according to Theorem 1. We suggested using a sensitivity analysis to address question (b), even if a case study does not tend to inform us of these probabilities with precision. And we demonstrated how to use the urn model under different assumptions about the evidentiary weight of observations, addressing (c). The need to address these questions is a limitation of our method. In Fisher's design, the randomization answered all three questions required to justify the urn model.

But can we still learn something similar to what we would have learned if we could randomize? To address this, let us assume that Fisher did not run a randomized experiment, instead he simply observed that Dr. Bristol correctly indicated the order of milk and tea after tasting 4 cups. If we apply our $+1$ hypergeometric model, we would have to assume a null distribution with 9 cups, including 5 theorized cups which Dr. Bristol would have guessed incorrectly. This distribution is consistent with the null but also with the 4 observations made. Strictly speaking, the only difference between our distribution and Fisher's own is that we would have to assume a total of 9 cups instead of 8. Making 4 correct guesses out of a total of 9 possible guesses ($p \leq 0.004$) is more rare than making 4 correct guesses out of 8 ($p = 0.014$). Substantially speaking, we arrive at the same conclusion: it would be very unlikely to observe a perfect run of 4 correct assessments if Dr. Bristol was indeed guessing. What if observations were biased? Say Dr. Bristol cheated. For bias to interfere with our substantial interpretation of results, bias would have to make correctly-tasted cups 2.6 times more likely than incorrectly-tasted cups before we could state that $p > .05$. Is this likely? Maybe yes if we have some reason to believe that the test was rigged in her favor.

---

[14]We say that our test emerges from design because it must include at least the number of observations supporting the working hypothesis that were actually made and because we do not otherwise assume a probability distribution for outcomes.

It is clear that our urn model would not turn a case study into a randomized experiment[15], and that the limitations of observational research remain. This is why we propose a sensitivity analysis as an integral part of the process.

# 9 Discussion and conclusion

The recent probabilistic turn in process tracing (Bennett 2008; Fairfield and Charman 2022; Humphreys and Jacobs 2015) renewed the method's focus on causal inference within cases (Falleti 2015; Collier 2011) and kicked off an important debate (Zaks 2021, 2022). We add a new approach to this discussion and expand the toolkit available to process tracers with a framework that builds on a key summary of evidence developed for randomized experiments: the *p*-value. Given that one cannot randomize history, we shifted the focus from randomization to conservativeness, or error avoidance, granted by the report of *p*-values' exact upper bound instead of Fisher's exact *p*.

Our test uses what we called a +1 hypergeometric null model to calculate a *p*-value upper bound and a measure of sensitivity expressed in odds-ratios. Together, the two probabilities allow us to cast doubt upon a rival theory about a single case knowing that a false positive is the least likely possibility within the model. Once a rival theory is rejected, the researcher can continue to run the test in other feasible rival explanations, indirectly accumulating confidence in her own theory. Our test does not compete with other uses of *p*-values in small-*N* research (e.g. Glynn and Ichino 2015), nor with other process tracing tests. Quite the opposite. Because it rests on few and simple assumptions about data, it allows us to incorporate *p*-values onto these earlier frameworks as a companion to process tracers following virtually any tradition.

We do, however, join Fairfield and Charman (2022) and others in opting for a probabilistic approach instead of process tracing's traditional reliance on ideas about necessity and sufficiency, borrowed from set theory (Beach and Pedersen 2019; Collier 2011; Van Evera 1997). For instance, to pass a smoking gun test researchers traditionally have to assume that an observation is impossible given the rival theory.[16] We find it easier to quantify the smoking gun quality of an observation via weighting and then reject a rival hypothesis by showing that observations are extremely unlikely, but still possible, in a null model that represents that rival. We find it useful to engage with probabilities regardless of the strength of evidence because one can always imagine an alternative causal driver that could have produced even the strongest smoking gun observation. The hoop test analogy is closer to what we propose, as it describes evidence that fits a theory but does not exclude a rival. But we ask how well it fits a rival instead. This is not to say that one should not use smoking gun or hoop testing frameworks. Researchers can integrate *p*-values into these tests by weighting the evidence, and we showed a simple framework for this as well as a graphical way to display how weighting affects results.

---

[15]Other limitations arise from the fact that we focus only on urns with 2 kinds of evidence rather than multiple kinds, and we limit the use of the procedure to tests of a single theory rather than multiple.

[16]A smoking gun test implies that a piece of evidence is sufficient for proving a theory.

In praise of a probabilistic approach, ours builds on previous contributions from scholars using Bayesian methods to calculate an easy to share sense of the level of confidence in working vs rival theories (Barrenechea and Mahoney 2019; Bennett 2008; Fairfield and Charman 2022; Humphreys and Jacobs 2015). These approaches ask researchers to input prior odds of observations and use Bayes rule to compare how theoretical priors would be updated in light of new information. Just as Fisherian hypothesis tests use less information and tend to be more conservative than Bayesian hypothesis tests, we too imagine that our approach is more conservative if compared to Bayesian alternatives.[17] But we must note here that calculating a *p*-value need not prevent one from using Bayesian methods. For example, we can imagine a Bayesian model that incorporates our urn-model in the likelihood term.

Neither does working with probabilities require that we discard the traditional inferential logic of process tracing. In effect, researchers have proposed different ways of combining set theory and probability reasoning (Barrenechea and Mahoney 2019). Ours is yet another possibility. In addition, we would like to think that the use of *p*-values can contribute for the development of clearer guidelines and replicability in those approaches too, a current demand in the qualitative methods literature (Zaks 2021, 2022).

Some may question if our approach is suited for comparative designs, which are within the scope of process tracing (Falleti 2015) and for which *p*-values have already been proposed (Glynn and Ichino 2015). We can certainly see uses of our model in research designs that account for multiple cases. For instance, the external validity of results could be entertained by replication in least or more likely cases (Beach and Pedersen 2019; Bennett and Elman 2007) or by matching cases using Mill's methods (Theda Skocpol and Margaret Somers 1980) and comparing *p*-values.

Our test takes inspiration from the logic behind a randomized experiment, even if no randomization is possible in a qualitative case study. But more skeptical researchers may question whether we need a measure such as a *p*-value in qualitative research in the first place, or if other standards for evaluation are better suited for within-case causal inference. Certainly we do not dismiss other criteria used in qualitative research, such as the idea of theoretical data saturation (Knott et al. 2022; Small 2009). As we see it, the information that our tests provides is different, rather than competing. By formalizing a model of observation we can address questions about how researchers observed what they did, relating to counterfactual scenarios that inform causality. By offering case-studies design a way to use a *p*-value we hope to encourage conversations across modes of research that enhance causal inference in the social sciences.

---

[17]One could certainly think of ways to simulate a comparison between the two approaches, even though there is not "one" Bayesian approach to process tracing but a few, and not always as formalized as our test in terms of how to assign probabilities. We defer such comparisons for other research.

# Table of contents

This document offers additional ways of calculating $p$-values following the +1 hypergeometric model for the testing a causal theory against a rival hypothesis, for instance when conducting process tracing. The purpose of our test is to facilitate the adoption of experimental reasoning (Morton and Williams 2010) in research dedicated to within-case causal inference. Here we demonstrate that the math behind our model works, offer an additional illustration with a real case study, Snows' classic cholera study, and present a beta version for an R package designed to run the +1 model with one simple line of code.

# A  Simulating the urn model

In the first simulation in our paper we portray a researcher with 2 observations from interviews, which are both favorable to her hypothesis. We then ask what is the probability of making the 2 observations supporting the working explanation $H_T$ if indeed the rival explanation $H_R$ governed the causal process? We showed that one can calculate this probability exactly using the model. Here we show how we can simulate drawing the urn using an algorithm, instead of directly calculating $p$. The algorithm is very simple. We posit an urn containing 2 items labeled "w" (for "working theory") and 3 items labeled "r" (for "rival theory"). We then draw 2 items and record whether or not both are labeled "w".

```r
## Using 1="working" and 0="rival"
urn <- c(rep(1, 2), rep(0, 3))
set.seed(12345)
res <- replicate(1000, sample(x = urn, size = 2, replace = FALSE))
num_working_balls <- apply(res, 2, sum)
kable(table(num_working_balls))
```

| num_working_balls | Freq |
|---|---|
| 0 | 289 |
| 1 | 602 |
| 2 | 109 |

```r
kable(table(num_working_balls) / 1000)
```

| num_working_balls | Freq |
|---|---|
| 0 | 0.289 |
| 1 | 0.602 |
| 2 | 0.109 |

```r
prop_working_balls <- mean(num_working_balls == 2)
num_working_balls_2 <- sum(num_working_balls == 2)
```

After 1000 draws, we get 109 w items, which amounts to 0.109 of the total. This is the same as saying that the probability of observing the 2 items supporting our working theory from this urn-model is 10.9%. Calculating $p$ directly using the hypergeometric distribution produced $p = .1$. We do not get the exact $p$ from this particular simulation simply because one run of 1000 simulations will produce different results from another run. As we increase the number of simulations the simulation-based *p*-value should become ever closer to the exact number. And, across simulations, the simulation-based calculations should not be systematically different from the exact calculation from the hypergeometric distribution (i.e. neither systematically too high or too low).

# B  Application in Snow's (1855) cholera study

We apply our test to Snow's (1855) famous cholera study, a classic in Epidemiology but also a constantly revisited reference in social science methodology (Dunning 2012) and, within it, process tracing (Freedman 20010; Collier 2011; Fairfield and Charman 2022). In 1854, the district of Soho in London experienced a cholera outburst. Snow theorized that cholera was caused by the ingestion of contaminated water, arguing against the prevailing explanation at the time: miasmatic theory. The latter explained that cholera and other illnesses were caused by "miasmas," believed to be present in bad air which itself originated spontaneously from filth.

Snow's key piece of evidence was observational: he produced a map of deaths by cholera in Soho. The map shows how fatalities were clustered around the pump of Broad Street, which Snow interpreted as evidence of the contamination of that source of water and its connection to the outburst. The map can be seen in Figure B.2.

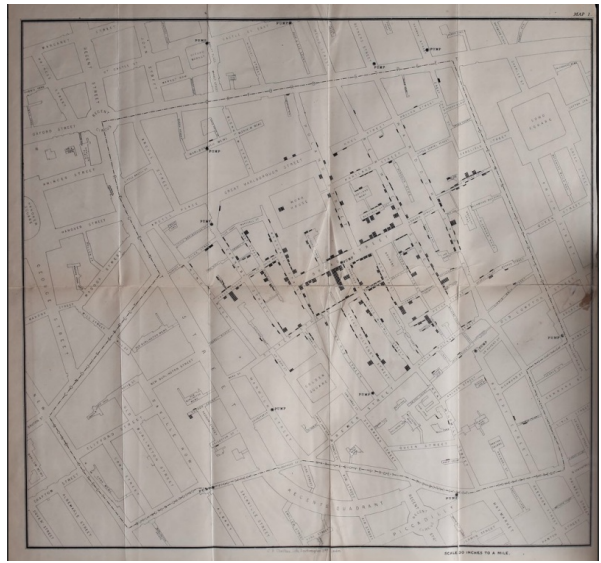

Figure B.2: Snow's (1855) map of cholera deaths in Soho

Access to the pump generated two groups, one group of individuals who lived in the premises of Broad Street and therefore relied on it for water (a group for which, the causal driver, $T$ is "use of the Broad Street pump" and has a value of $T = 1$), and a another group of individuals who were too far from the pump for it to be a convenient source of water (people putatively not exposed to water from the Broad Street pump, $T = 0$). Snow could have calculated Fisher's exact *p*-value directly under the assumption that this was a natural experiment with "as-if random" treatment assignment (Dunning 2012). However, Snow addressed this as a case study where a theorized cause is present and an outcome is observed. The visual correlation between deaths and the pump, and the reasoning about each observation that we describe below, supported the causal theory connecting the water from the pump at Broad Street to death by cholera.

To infer the causal effect of drinking water from the Broad Street pump, Snow implemented what we would now call process tracing (Freedman 20010; Collier 2011). He conducted interviews and archival research to describe the sequence of events in Soho and its surroundings while these deaths were occurring (Snow 1855, 42–44). Snow interviewed at least four individuals and got access to official records, all of which provided a series of relevant observations connecting deaths by cholera to the ingestion of water from the Broad Street pump.

Snow's findings granted him a place in the history of Epidemiology and Microbiology. The observational evidence did strongly support the contaminated water theory and it foreshadowed the discovery of *Vibrio cholerae*, the bacteria that transmits cholera through human stool. Eventually, evidence against miasmatic theory became overwhelming, and all subsequent research indicates that miasmas do not exist. However, at the time of publication Snow's study was received with skepticism by many and belief in miasmatic theory remained. This occurred because, contrary to our intuition today, miasmatic theory was a strong rival to the idea of water borne contamination. Belief in miasmas was supported by the fact that miasmatic theory allowed for correct predictions. For instance, efforts to get rid of bad air by increasing ventilation and incentivizing hygiene indeed reduced the spreading of disease. Assuming that we do not know of the existence of microorganisms, how strong was Snow's evidence against the miasmatic theory rival? Let us list his observations below.

*Observation 1*: The map. As noted, Snow mapped all deaths in Soho and found a striking coincidence between deaths and proximity to the pump at Broad Street. Today, one could easily summarize the geo-referenced information as a dataset on its own (say, by describing survival as a function of distance to the pump). However, let us take this piece of evidence as it was at the time: as a single document showing the distribution of deaths.[18]

*Observation 2*: Snow learned that of the 537 inmates from a workhouse in Poland Street, right next to the pump at Broad Street, only five died of cholera. The workhouse had its own well and water from Broad Street was therefore rarely consumed.

*Observation 3*: Snow was informed about another case of apparent immunity: a brewery near the Broad Street pump in which none of the seventy plus workers fell ill. The owner of the brewery informed Snow that workers mostly drank beer and also had their own well if they felt the need of water.

---

[18] After all, a statistical survival analysis would also boil down to one piece of information about the map. Snow just summarized the map evidence using an implicit mental model of his own.

*Observation 4*: In contrast to the workhouse and the brewery, a nearby percussion-cap factory was equipped with pipes that channeled water directly from the Broad Street pump and eighteen of its workers died of cholera.

*Observation 5*: Next to the brewery, seven workmen in a factory of dentists' materials also drank from the Broad Street and died.

*Observation 6*: Two other individuals who were also neighbors to the percussion-cap factory were not in the habit of drinking water from Broad Street and they survived.

*Observation 7*: Snow also learned that an officer of the army came to dine in nearby Wardour Street where he was served water from the Broad Street pump and died within a few hours.

*Observation 8*: In addition, a young pregnant woman on Bentinck Street who decided to go to Broad Street for water, which was unusual for her, passed away the following day.

*Observation 9*: An interviewee also informed Snow that a gentleman in delicate health went to visit his ill brother in 6 Poland Street. The visitor ingested Broad Street water with brandy, dying on the evening of the following day. The ill brother also died of cholera.

*Observation 10*: Snow was also informed about the widow of a percussion-cap maker in the Hampstead district (far from Broad Street) who was "attacked" by cholera but had not been near the contaminated pump of Broad Street. Her niece, who had recently visited her, also died of cholera. The widow's son informed Snow that a cart regularly delivered a bottle of water from Broad Street to her mother "who preferred it."

Snow considered that these 10 observations provided sufficient evidence about the mechanisms that accounted for survivors living near the pump and infections far from the pump, thus supporting his theory. The observations accounted for different explanations for why the water did not infect individuals in the treatment group (having their own well and hydration by beer) and why cholera reached the lady outside the pump radius (preferring contaminated water). The observations are compelling indeed, but they do not constitute a test by themselves.

For a hypothesis test, we must first articulate or identify a rival theory to generate a rival hypothesis. In Snow's case, miasmatic theory plays the role of the rival, $H_R$. If miasmatic theory is correct, then the pump had nothing to do with the observed deaths. The second step is describing how the observations could support miasmatic theory. Recall that all observations must be possible under the null hypothesis $H_R$. The job of the hypothesis test is to summarize how probable it would be to make the observations categorized as supporting the water theory while entertaining, for the sake of argument, the rival theory.

Focusing on Snow's most famous piece of evidence, his map (observation 1), one would think that bad air should not frequently be found around a water pump. However, bad air could be condensed there because of meteorological phenomena that are unrelated to the pump. Thinking of Soho independently of the pump, the prevalence of deaths by cholera there is not unexpected under the rival theory because Soho was a badly sanitized industrial district, and filth produces miasmas, according to the theory $H_R$. It would not be odd to find miasmas near the pump considering the filth from, say, the poor house and all the factories right next to it. Therefore, we categorize observation 1 as supporting miasmatic theory. In effect, because of the bad air of Soho, we can code all deaths in its surroundings as expected in the hypothetical world of the rival explanation (observations 4 and 5). On the other hand, observations of dispersed cholera infections of individuals that were not usually exposed to the bad air of Soho should not support the rival (observations 7, 8, 9, and 10). Likewise, the cases of absence of cholera within the epicenter of the disease should not support the rival explanation (observations 2, 3, 6, and 10). However unlikely, these observations could still happen in the world of the rival, as maybe a gust of wind carried miasmas afar saving some and killing others.

From the 10 observations in Snows' study, we would categorize 3 as members of $\mathcal{R}$ because they are somewhat expected under the rival explanation. Meanwhile, 7 observations (the ones more clearly supporting his theory) would have to belong to set $\mathcal{T}$ because they would be very odd (but not impossible) if the null was true. To entertain the rival, we assume that $|\mathcal{R}| > |\mathcal{T}|$. To produce the most conservative test, $\mathcal{R}$ has to have $7 + 1 = 8$ items and so that urn must have a total of 15 items. Table B.4 shows the probability of observing each possible number of rival-theory supporting observations given this urn model.

Table B.4: Probabilities of observing 0–10 observations against the rival hypothesis

| # Obs. | 0 | 1 | 2 | 3 | 4 | 5 | 6 | 7 | 8 | 9 | 10 |
|---|---|---|---|---|---|---|---|---|---|---|---|
| Prob. | 0.00 | 0.00 | 0.01 | 0.09 | 0.33 | 0.39 | 0.16 | 0.02 | 0.00 | 0.00 | 0.00 |

The probabilities in Table B.4 speak to theorized repetitions of Snow's study, each time gathering a different amount of observations against the Rival. We know that in reality Snow made 7 such observations. Therefore, the information that one wants is the probability of observing such data, or even more favorable data regarding his preferred theory of water born illness. We summarize this information along with the proposed sensitivity analysis in Table B.5.

Table B.5: Test summary for Snow (1855)

| $p$-value upper bound | Odds ratio for $\omega_{\alpha=.05}$ | Odds ratio for $\omega_{\alpha=.10}$ |
|---|---|---|
| 0.019 | 1.59 | 2.36 |

In the hypothetical world of miasmatic theory one would have observed what Snow observed (7 observations against the miasma explanation) only about 2% of the time or less ($p \leq 0.019$) if all of the pump-explanation supporting observations and miasma-explanation supporting observations were equally likely to be made or equally easy to be made. The odds-ratio needed to invalidate results at a $p=0.05$ level is 1.59. That is, pump-supporting evidence would have to be 60% more likely (or easier) to be made for the *p*-value to rise above $p = .05$. To rise above the threshold of $p = .10$, bias, from whatever source, would have to make account for an odds-ratio of 2.36 in favor of pump-supporting evidence. This means that an argument against Snow's results would need to explain how it was that Snow's observation choices were more than twice as likely to support the working explanation than the rival. In the case of Snow's careful research design — in which he went out of his way to seek observations that would contradict his pump theory — assuming this amount of bias seems unreasonable to us. Based on the odds-ratios needed to flip conclusions, one can argue with confidence that the sensitivity of this test to bias is very low. Of course, a miasma expert could respond by listing the many other observations that Snow did not record that would have supported the rival theory. If that work showed that, in fact, for each piece of pump-supporting information there were roughly two other pieces of miasma-supporting observations, then this expert could produce an argument against Snow's conclusions. In the absence of such an engagement with the evidence base, however, we end our re-analysis of Snow's data concluding that the argument against miasma theory is strong and ought to guide future public policy regarding cholera.

## C  Explanation of why the Hypergeometric Distribution describes an Urn Model

The urn model used in this paper is one possible urn model among many urn-models. We build on Fisher's model for this paper because it connects most directly with the situation where a researcher can categorize pieces of evidence as supporting a focal theory or not. The model portrays a relatively small number of pieces of discrete evidence inside the urn and a process of drawing from the urn which does not replace the items after each selection. The probability mass function for the urn model gives the probability that the researcher would draw $x$ observations from $|\mathcal{T}|$ (in favor of the working theory) given a draw of size $n$ from the urn, assuming that the urn contains both $\mathcal{T}$ and $\mathcal{R}$, and that $|\mathcal{R}| = |\mathcal{T}| + 1$. If the we write the total number of pieces of informationg in the urn as $U$, we can write $U = |\mathcal{R}| + |\mathcal{T}| + 1 = (2|\mathcal{T}|) + 1$. The mathematical expression for this probability is:

$$P(X = x) = \frac{\binom{|\mathcal{T}|}{x}\binom{U-|\mathcal{T}|}{n-x}}{\binom{U}{n}}. \tag{C.2}$$

Here, we explain how this mathematical formula arises as a description of the model. We start with an urn containing $U = 5$ pieces of information in total, divided into $|\mathcal{T}| = 2$ pieces that support the focal theory and $U - |\mathcal{T}| = 5 - 2 = 3$ pieces that support the rival theory. In this model, the researcher chooses $n = 2$ items at a time from this urn. And we want to know the probability of observing $x = 2$ draws from $\mathcal{T}$. The model assumes that each piece of information has an equal probability of being drawn and that after 5 draws all pieces have been drawn and thus we see the whole contents of the urn. The shorthand calculation is of this probability is: $P(X = 2) = \frac{\binom{2}{2}\binom{5-2}{2-2}}{\binom{5}{2}} = \frac{1}{10}$.

Why does this calculation make sense? Our question about probability can be restated in terms of events: how many ways can one draw 5 items from the model? Out of those ways, how many would involve drawing 2 pieces of evidence favoring the working or maintained theory? In Figure C.3 we show the 120 possibilities for drawing 5 pieces of information. With no loss of generality, we label pieces of information "1" and "2" as supporting the working theory and the other pieces as supporting the rival. The figure shows that 12 total draws from the urn involve the pieces of information labeled "1" and "2" (they are the six draws that start with the "1" piece followed by "2" — $(1, 2, 3, 4, 5), (1, 2, 4, 3, 5), \ldots$ — and the six that start with "2" and follow with "1" — $(2, 1, 3, 4, 5), (2, 1, 4, 3, 5), \ldots$). So, what is the probability of drawing 2 pieces of evidence favoring the working theory in an urn-model designed to favor the rival theory? It is 12/120 or 1/10.

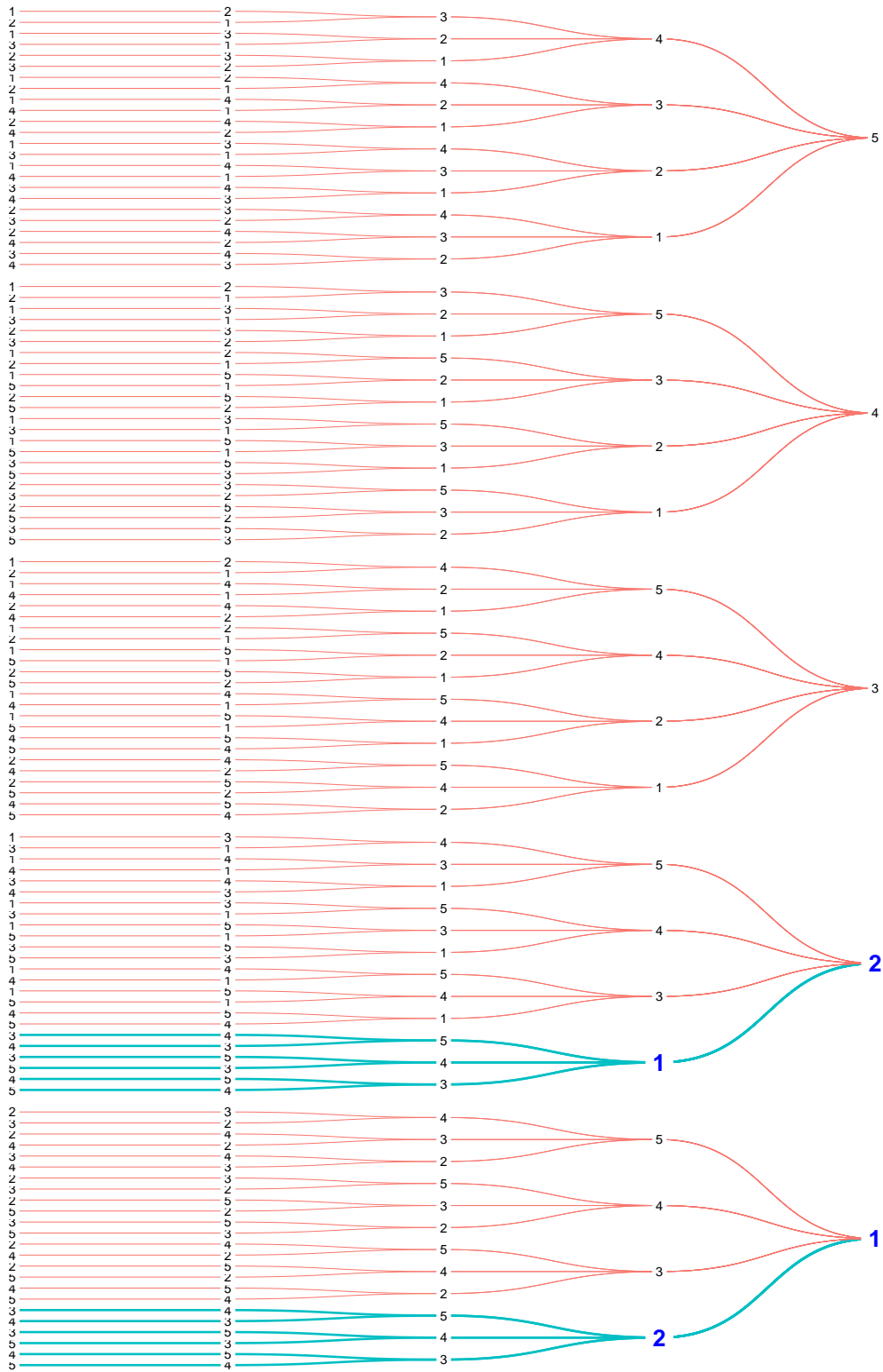

Figure C.3: Possible draws without replacement from an urn with 2 pieces of information supporting the working theory (if first 2 draws, in blue, bold font) and 3 pieces of information supporting the rival (in black).

The formula in C.2 can be broken up into parts. Throughout it uses the binomial coefficient notation of $\binom{a}{b}$ which is short hand for "the number of ways to choose $b$ items from a set containing $a$ items" and can be written using multiplication as $\frac{a!}{b!(a-b)!}$. In our case, for example, if we want to count the total number of ways to choose 2 items from a set of 5, we could write $\frac{5!}{2!(5-2)!} = \frac{5 \cdot 4 \cdot 3 \cdot 2 \cdot 1}{(2 \cdot 1)(3 \cdot 2 \cdot 1)} = \frac{120}{12} = 10$. An we take the chance here to highlight that the hypergeometric model converges to a binomial distribution

Why do we use this multiplication to count the number of ways we could draw 5 items from an urn without replacement? Why do we use the multiplication $5! = 5 \cdot 4 \cdot 3 \cdot 2 \cdot 1$? The figure highlights that on the first draw, one can choose any of the five items (1,2,3,4 or 5), but once a first item has been drawn there are only 4 items left. For example, if the first draw is item "1", then the second draw from that urn can only be items 2,3,4, or 5. So there are 5 ways to draw the first item, and for each of the 5 ways to draw the first item, there are 4 ways to draw the second item: $5 \cdot 4$. After the second item, there are only 3 items left. So, there are 3 ways to choose a third item for each of the 4 ways to choose a second item for each of the 5 ways to choose a first item: $5 \cdot 4 \cdot 3$. The figure shows this expansion of possibilities graphically.

The numerator of C.2 counts the number of ways one can draw $k$ items from the total of items supporting the working hypothesis, $|\mathcal{T}|$ written $\binom{|\mathcal{T}|}{x}$. At the moment, the urn contains exactly 2 such items, and we are drawing then 2 at a time, so $\binom{2}{2} = 1$. The numerator is the number of ways to choose 2 items from the set of observations that support the rival: $\binom{U-|\mathcal{T}|}{n-x}$. In our case there are 3 such observations so we have $\binom{3}{0}$ (because we are drawing items 2 at a time from a set of 2), and the number of ways to draw 0 items from 3 is 1, so we have 1. The numerator of the formula is thus 1 and the denominator is 10.

One can verify these calculations using the `dhyper` command that is built into R, or the `FisherHypergeometricDistribution` command in Mathematica, the `hypergeometric` command in Stata, or the `hypergeom.sf` command from `scipy.stats` in python. We also asked ChatGPT this question and got the right answer.

## D Computing for Sensitivity Analysis

We showed that we could find the value for $\omega$ at which the $p$-value for our observed evidence equalled .05 using R with the following code. Here we also show how we used Mathematica to confirm our results. Here we are using $n = 3$, $|\mathcal{T}| = 2$, and $|\mathcal{R}| = 3$.

```
find_odds <- function(omega, alpha_thresh = .05) {
  # m1= number of pieces of evidence supporting the working theory
  # m2 = number of pieces of evidence supporting the rival theory
  # n = number of pieces drawn from the urn
  # odds = odds of drawing evidence supporting the working theory versus
  ↪   rival theory
```

```
  p_found <- dFNCHypergeo(x = 2, m1 = 2, m2 = 3, n = 3, odds = omega)
  return(p_found - alpha_thresh)
}

found_odds <- uniroot(f = find_odds, interval = c(0, 10))$root
found_odds
```

[1] 0.195

Here we show both the input and output of the Mathematica commands:

```
In[1]:= theeq = (3*w^2)/(1 + 6*w + 3*w^2) == p
Out[1]:= (3 w^2)/(1 + 6 w + 3 w^2) == p
In[2]:= FindRoot[theeq /. p -> .05, {w, 1, .00001, Infinity}]
Out[2]:= {w -> 0.195159}
In[3]:= NSolve[(theeq /. p -> .05) && w > 0, w, Reals]
Out[3]:= {{w -> 0.195159}}
```

We get the same answer if we use Wolfram Alpha's free online math service ( at https://wolframalpha.com) by asking: "Please find the value of w for which $(3w^2)/(1 + 6w + 3*w^2) == .05$ where w>0."

The numerical approach shows that we did not need to solve for $\omega$. If we did want to do so, for a particular choice of $x$, $|\mathcal{T}|$, $|\mathcal{R}|$, and $n$, we could again use Mathematica to check our work. We note that once we have more than 3 balls in the urn the solutions for higher-order polynomials become more difficult to use. So, for general purpose use, we recommend the numerical approach.

```
In[4]:= Solve[theeq, w, Reals, Assumptions -> w > 0 && 0 < p < 1]
Out[4]:= {{w -> -(p/(-1 + p)) + Sqrt[p + 2 p^2]/(Sqrt[3] (1 - p))}}
{{w -> -(p/(-1 + p)) + Sqrt[p + 2 p\^2$$/(Sqrt[3] (1 - p))}}
```

The online WolframAlpha system also can solve this equation if we ask "Please solve `(3*w^2)/(1 + 6*w + 3*w^2) == p` for w where w>0 and 0<p<1".

We can also report that ChatGPT 4o produces the same solution for $\omega$ in terms of $p$ for a given set of $x$, $|\mathcal{T}|$, $|\mathcal{R}|$, and $n$.

# E  Proof of Theorem 1 on Conservativeness of the +1 Urn Model

Although we know how many observations support the working theory, $|\mathcal{T}|$, we want to know how many observations should support the rival theory in the urn model. We show here that the urn containing $|\mathcal{R}| = |\mathcal{T}| + 1$ is the version of the urn model that gives the most benefit of the doubt to the rival theory, and thus, is the urn model we advocate for use in assessing whether $\mathcal{T}$ contains sufficient evidence against the rival. Below, we operationalize "most conservative" or "most benefit of the doubt to the rival" by "largest probability of seeing $x$ or more working theory supporting observations" or "largest $p$-value".

Although we tend to write the total size of the urn model as $U$, for the purposes of the theorem we replace $U = |\mathcal{R}| + |\mathcal{T}|$ since the urn must contain pieces of evidence supporting those two theories. Also, since we want to know how to set $|\mathcal{R}|$ compared to $|\mathcal{T}|$, we will write $|\mathcal{R}| = |\mathcal{T}| + c$ and focus on the question of finding the value for $c$ at which the probability of seeing an observation from $\mathcal{T}$ is highest. This means that we will write the probability mass function (pmf) as

$$P(x) = \frac{\binom{|\mathcal{T}|}{x}\binom{U-|\mathcal{T}|}{n-x}}{\binom{U}{n}} == \frac{\binom{|\mathcal{T}|}{x}\binom{(|\mathcal{R}|+|\mathcal{T}|)-|\mathcal{T}|}{n-x}}{\binom{|\mathcal{R}|+|\mathcal{T}|}{n}}. \tag{E.3}$$

And this simplifies after substituting $|\mathcal{R}| = |\mathcal{T}| + c$ to:

$$P(x) = \frac{\binom{|\mathcal{T}|}{x}\binom{|\mathcal{T}|+c}{n-x}}{\binom{|\mathcal{T}|+c+|\mathcal{T}|}{n}} = \frac{\binom{|\mathcal{T}|}{x}\binom{|\mathcal{T}|+c}{n-x}}{\binom{2|\mathcal{T}|+c}{n}}. \tag{E.4}$$

To simplify the reading experience below we write $T = |\mathcal{T}|$.

**Theorem 2.** *Consider the probability mass function of a hypergeometric distribution written as a function of $T$.*

$$P(x) = \frac{\binom{T}{x}\binom{T+c}{n-x}}{\binom{2T+c}{n}}, \tag{E.5}$$

*where $0 \leq x \leq n \leq T$, $n = T$, and $T, n, x, c$ are nonnegative integers. The probability of seeing $x$ or more observations supporting $\mathcal{T}$ is calculated using the tail probability of the cumulative density function of this distribution:*

$$F(x) = P(X \leq x) = \sum_{j=0}^{x} \frac{\binom{T}{j}\binom{T+c}{n-j}}{\binom{2T+c}{n}}. \tag{E.6}$$

*This theorem states that, for fixed $x, n, T$ the tail probability, $1 - F(x) = P(X > x) = \sum_{j=x+1}^{n} \frac{\binom{T}{j}\binom{T+c}{n-j}}{\binom{2T+c}{n}}$, is maximized when $c = 1$.*

*Proof.* Consider the tail probability:

$$1 - F(x) = P(X > x) = \sum_{j=x+1}^{n} \frac{\binom{T}{j}\binom{T+c}{n-j}}{\binom{2T+c}{n}}. \tag{E.7}$$

1. We see that increasing $c$ adds more rival supporting observations to the urn (recall that $T + c$ are the number of observations supporting the rival theory), which increases the total number of ways to choose $n$ sampled observations, so the denominator, $\binom{2T+c}{n}$, becomes larger with $c$.

2. The numerator for each term in the sum, $\binom{T}{j}\binom{T+c}{n-j}$, also grows as $c$ increases, but not as fast as the denominator $\binom{2T+c}{n}$.

3. Since $\binom{T}{j}$ is constant in $c$, we can compare $\binom{T+c}{n-j}$ with $\binom{2T+c}{n}$ to see this difference in growth rate. Here $\binom{T+c}{n-j}$ grows proportionally to $(T+c)!$ and $\binom{2T+c}{n}$ grows proportionally to $(2T+c)!$. Notice that neither $n$ nor $n - j$ depend on $c$. Since $2T + x >= T + x$, we have: $(2T + x)! > (T + x)!$.

4. So, the denominator's growth outpaces the numerator's growth as $c$ becomes large, because adding rival supporting observations disproportionately increases the number of possible combinations of $n$ observations in total. We can see this difference in rates of change of the numerator versus the denominator in Figure E.4:

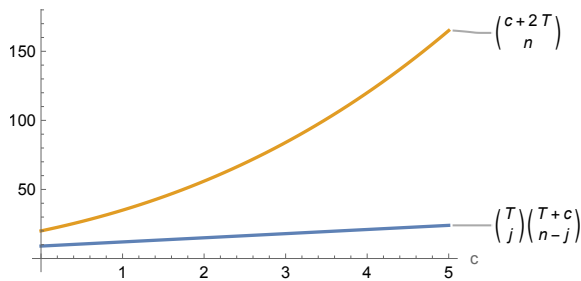

Figure E.4: A smoothed plot of the values of the numerator versus the denominator values as $c$ increases.

5. Thus, as $c$ increases, each term $\frac{\binom{T}{j}\binom{T+c}{n-j}}{\binom{2T+c}{n}}$ decreases. Summing over $j = x + 1$ to $n$, we find that $P(X > x) = 1 - F(x)$ decreases as $c$ increases.

Since $c \geq 1$, the maximum of $1 - F(x)$ occurs at the smallest possible $c$, which is $c = 1$. □

# F Introducing the R package DrBristol

Here we introduce a beta version of the R package DrBristol, named after the "lady tasting tea" in Fisher's experiment. The package allows users to calculate the upper bound of *p*-values when testing theories about single cases, as well as a measure of sensitivity. The R package `DrBristol` can be installed from Github using the `remotes` package and the command `install_github()`.[19] The following code demonstrates the use of the functions in that package used to create *p*-values under different scenarios.[20]

```
library(DrBristol)
# Equal probability, 2 kinds of evidence
find_p_two_types(obs_support = 7, total_obs = 10)
```

```
[1] 0.0186
```

```
# Equal probability, 2 kinds of evidence with interpretation printed
find_p_two_types(obs_support = 7, total_obs = 10, interpretation = TRUE)
```

```
The maximum probability of drawing 7 observations which support the working
theory from an urn model supporting a rival theory, where the odds of
observing working theory information is odds=1 and evidentiary weights are
(1,1,1,1,1,1,1), is p <=0.0186.

$thep
[1] 0.0186

$interp
[1] "The maximum probability of drawing 7 observations which support the
working theory from an urn model supporting a rival theory, where the odds of
observing working theory information is odds=1 and evidentiary weights are
(1,1,1,1,1,1,1), is p <=0.0186."
```

```
# Unequal probability, 2 kinds of evidence with interpretation printed
find_p_two_types(obs_support = 7, total_obs = 10, interpretation = TRUE, odds
↪    = .5)
```

```
The maximum probability of drawing 7 observations which support the working
theory from an urn model supporting a rival theory, where the odds of
observing working theory information is odds=0.5 and evidentiary weights are
(1,1,1,1,1,1,1), is p <=0.003.
```

---

[19]Github repository is currently anonymized.

[20]The current development of the method is limited to two kinds of evidence but we anticipate extending the method to multiple kinds and multiple theories.

```
$thep
[1] 0.00299

$interp
[1] "The maximum probability of drawing 7 observations which support the
working theory from an urn model supporting a rival theory, where the odds of
observing working theory information is odds=0.5 and evidentiary weights are
(1,1,1,1,1,1,1), is p <=0.003."
```

```r
find_p_two_types(obs_support = 7, total_obs = 10, interpretation = TRUE, odds
  = 2)
```

The maximum probability of drawing 7 observations which support the working
theory from an urn model supporting a rival theory, where the odds of
observing working theory information is odds=2 and evidentiary weights are
(1,1,1,1,1,1,1), is p <=0.0761.

```
$thep
[1] 0.0761

$interp
[1] "The maximum probability of drawing 7 observations which support the
working theory from an urn model supporting a rival theory, where the odds of
observing working theory information is odds=2 and evidentiary weights are
(1,1,1,1,1,1,1), is p <=0.0761."
```

```r
# Equal probability, Unequal evidentiary weight, 2 kinds of evidence
find_p_two_types(
  obs_support = 7, total_obs = 10, weights = rep(1, 7),
  interpretation = TRUE, odds = 1
)
```

The maximum probability of drawing 7 observations which support the working
theory from an urn model supporting a rival theory, where the odds of
observing working theory information is odds=1 and evidentiary weights are
(1,1,1,1,1,1,1), is p <=0.0186.

```
$thep
[1] 0.0186

$interp
```

[1] "The maximum probability of drawing 7 observations which support the
working theory from an urn model supporting a rival theory, where the odds of
observing working theory information is odds=1 and evidentiary weights are
(1,1,1,1,1,1,1), is p <=0.0186."

```
find_p_two_types(
  obs_support = 7, total_obs = 10,
  weights = rep(c(2, 1), c(1, 7 - 1)), interpretation = TRUE, odds = 1
)
```

The maximum probability of drawing 7 observations which support the working
theory from an urn model supporting a rival theory, where the odds of
observing working theory information is odds=1 and evidentiary weights are
(2,1,1,1,1,1,1), is p <=0.0105.

$thep
[1] 0.0105

$interp
[1] "The maximum probability of drawing 7 observations which support the
working theory from an urn model supporting a rival theory, where the odds of
observing working theory information is odds=1 and evidentiary weights are
(2,1,1,1,1,1,1), is p <=0.0105."

This code shows how to use `DrBristol` to execute the sensitivity analyses that we developed in this paper.

```
# What is the odds that would bring our p=.02 to p \approx .05
find_p_two_types(obs_support = 7, total_obs = 10, odds = 1)
```

[1] 0.0186

```
the_odds <- sens_urn(obs_support = 7, total_obs = 10, p_threshold = .05)$w
print(the_odds)
```

[1] 1.59

```
# Notice that when we plug that odds into the p-value function, we get a
# value very near the p-value threshold
find_p_two_types(obs_support = 7, total_obs = 10, odds = the_odds)
```

[1] 0.05

# G References